\documentclass[5p,times]{elsarticle}

\usepackage{listings}

\providecommand{\Tsys}{\ensuremath{T_{sys}}}
\usepackage{lineno,hyperref}
\modulolinenumbers[5]
\usepackage{enumitem}

\journal{Astronomy and Computation}
\usepackage[figuresright]{rotating}
\usepackage{color}

\def\mysingleq#1{`#1'}

\begin{document}


\begin{frontmatter}
\title{The Data Processing Pipeline for the Herschel\footnote{\it{Herschel} \rm is an ESA space observatory with science instruments provided by European-led Principal Investigator consortia and 
with important participation from NASA.} - HIFI Instrument}

\author{K. Edwards}
\address{University of Waterloo, Waterloo, Canada}

\author{R.F. Shipman}
\address{SRON Netherlands Institute for Space Research, The Netherlands}

\author{D. Kester}
\address{SRON Netherlands Institute for Space Research, The Netherlands}

\author{A. Lorenzani}
\address{INAF Osservatorio Astrofisico di Arcetri, Florence, Italy}

\author{M. Melchior }
\address{Fachhochschule Nordwestschweiz, Switzerland}

\date{}

\begin{abstract}

The HIFI data processing pipeline was developed to systematically process diagnostic, 
calibration and astronomical observations taken with the HIFI science instrument
 as part of the Herschel mission. 
 The HIFI pipeline processed data from all HIFI observing modes within the Herschel 
automated processing environment, as well as, within an interactive environment.
 A common software framework was developed to best support the use cases required by the instrument teams and by the general astronomers. 
 The HIFI pipeline was built on top of that and was designed with a high degree of modularity. 
 This modular design provided the necessary flexibility and extensibility to deal with the complexity of batch-processing eighteen different observing modes, to support the astronomers in the interactive analysis and to cope with adjustments necessary to improve the pipeline and the quality of the end-products. 
This approach to the software development and data processing effort was arrived at by coalescing the lessons learned from similar research based projects with the understanding that a degree of foresight was required given the overall length of the project.   
In this article, both the successes and challenges of the HIFI software development process are presented.  To support future similar projects and retain experience gained lessons learned are extracted.

\end{abstract}

\begin{keyword}
{Methods: data analysis -- Physical sciences and engineering: astronomy -- Techniques: spectroscopic}
\end{keyword}

\end{frontmatter}

\section{Introduction}
\label{sec:intro}
The Herschel Space Observatory \citep{pilbratt2010} performed astronomical observations in the far infra-red and sub-millimeter  and  was comprised of three science instruments: the Heterodyne Instrument for Far Infrared (HIFI) \citep{degraauw2010} , the Photodetector Array Camera and Spectrometer (PACS) \citep{PACS}  and the Spectral and Photometric Imaging Receiver (SPIRE) \citep{SPIRE}. The Herschel Space Observatory was launched on May 14, 2009 from the Guiana Space Centre in French Guiana. The telescope and instruments ceased science operations on April 29, 2013 after the liquid helium cryogen used to cool the instruments had been exhausted.

HIFI was a heterodyne spectrometer operating between 480 GHz to nearly 2 THz. 
The HIFI instrument hardware and resulting capabilities are described in \citep{degraauw2010}.   
Extensive online HIFI documentation can be found in the Herschel Explanatory Legacy Library\footnote{\href{https://www.cosmos.esa.int/web/herschel/legacy-documentation-hifi}{https://www.cosmos.esa.int/web/herschel/legacy-documentation-hifi}\label{hell_fn}} where a thorough overview of the HIFI instrument is documented in the HIFI Handbook\citep{HIFIH}.  

Over the course of the Herschel mission, the HIFI instrument performed over 9100 scientific and calibration observations.  All observations taken with the HIFI instrument (including preflight test data) were passed through a series of standard processing steps which made up the HIFI data processing pipeline \citep{shipman2017}. 
The resulting observation products are publicly available from the Herschel Science Archive\footnote{\href{https://www.cosmos.esa.int/web/herschel/science-archive}{https://www.cosmos.esa.int/web/herschel/science-archive}\label{hsa_fn}} (HSA).

The main requirements for the data processing software was to correct for and remove instrumental artifacts in the observational data in an efficient and robust manner  while generating data products that could be used to answer fundamental astrophysical questions.  While this was the general aim of the data processing software and the Herschel mission itself, in order to achieve this goal, a comprehensive software framework was required. 
%
%
The implementation of this goal was driven by the collective experience of individuals from previous space missions as well as from other systems and software engineering projects {\citep{2003sws..bookR....L, 2003ESASP.481..387L, 2003ESASP.481..285W}}. 
 Accordingly, the development of the data processing software for the overall mission was incorporated into the planning of Herschel at a much earlier stage than had been done in previous missions. The purpose of this article is to describe the resulting software patterns, data structures and tools developed for HIFI that resulted from the acquired knowledge of how to best represent and analyze the HIFI instrument data for scientific gains.

This article is structured as follows: Section \ref{sec:hcss} gives details on the software development process for the Herschel mission and how the software framework was designed to support three very different instrument pipelines. 
Section \ref{sec:structures} describes the data and data structures used to organize the HIFI observational data in an efficient manner. 
Section \ref{sec:software} focuses on the HIFI pipeline and covers the full range of processing that was needed for HIFI data from standard product generation to quality assessment.  
Section \ref{sec:discussion} covers items of critical importance to the development effort which require further reflection as well as lessons learned.  Our main conclusions are presented in section \ref{sec:summary}.

\section{Herschel Common Software System (HCSS)}
\label{sec:hcss}

It was realized early on in the planning stages of the mission that despite the differences in each science instrument, there was quite a bit of common ground between the instruments and the observatory with regards to the data processing software requirements. The needs of instrument scientists, engineers and astronomers also had to be supported over the different phases of the mission. These common functional requirements ranged from having numerics libraries, a pipeline processing environment and archive query functionality. In order to generate high-quality data products and serve them through a common data archive, sound software development procedures were necessary to ensure production-ready reliable software.  In an effort to address these common requirements, the Herschel Common Software System (HCSS)\footnote{HCSS was a joint development by the Herschel Science Ground Segment Consortium, consisting of ESA, the NASA Herschel Science Center, and the HIFI, PACS and SPIRE consortia} was envisioned {\cite{bauer1998, riedinger2009}}. The HCSS development was coordinated by observatory staff (ESA) developers.  Instrument team developers were responsible for developing their own instrument specific software.  Together, the observatory and instrument developers were responsible for developing the common software infrastructure.

Management of the effort was important as the software development was done in parallel with the hardware development of the observatory and the instruments. The software tools were of critical importance in order to test the hardware in the laboratory before launch. Major portions of the data processing infrastructure had to be completed before launch as the Herschel mission had a limited lifetime related to the liquid helium supply. The software was needed in order to monitor the observatory and the instruments so as to not lose valuable observing time.

The software development process for the Herschel mission was lengthy and complex, spanning nearly two decades with a distributed group of developers primarily across Europe and North America. Early software design decisions made within the software project had lasting, and long term consequences for the success of the development project. One of the early decisions taken was the selection of a set of software languages to use for implementation, see section \ref{sec:softwareLanguages}. Another was a focus on software testing to ensure the correctness, robustness and consistency of the software which led to significant effort being placed on the development of unit, system and later astronomer acceptance testing, see section \ref{sec:configControl}.
The software developed for the Herschel mission followed the principles of an agile software development project, including cross-functional teams, continuous integration and iterative and incremental development. All changes introduced into the development builds were incremental in nature and were available to all users immediately after compilation and the successful execution of the unit tests. Breaking changes to the software were not permitted and backwards compatibility was expected between each major release along with deprecation warnings allowing developers and users to make timely changes.

\subsection{Data Processing {\it Modes}}
\label{sec:dataSoftware}

Herschel data processing software was required to support multiple {\it modes of operation} across the different mission phases. The major modes of operation consisted of {\it preflight} (in-lab event driven hardware testing), {\it systematic} (operations - batch processing - multiple instruments), {\it on-demand} (HSA - public reprocessing requests) and {\it interactive} (data analysis). 
The systematic mode was encapsulated in a software  framework referred to as the Standard Product Generation (SPG), see section  \ref{sec:SPG}.

During instrument level testing (ILT) in the initial phase of the mission, data processing software tools were required to process data generated from the pre-flight and flight hardware. These analysis tools, referred to as Quick Look Analysis (QLA) tools, were developed with a focus on being event driven as the data was streamed from the instruments in real time. 
It was necessary to have a (near) real-time overview of the instrument behaviour as inputs were being varied during testing. 
It was during the ILT phase of Herschel that many of the main algorithms used in data processing pipelines for the specific instrument analysis and data calibration were defined, developed and tested. These algorithms were encapsulated as tasks, see section \ref{sec:taskDescription}, and formed the basis of the instrument pipelines. Re-using software developed in this phase, greatly enhanced the reliability of those tools later in the operations phase. 

In the operations phase of the mission the need for real-time analysis diminished as ground communications with the observatory were limited. In operations, the observatory and instruments were designed to autonomously collect data for about eighteen hours a day.  There was a nominal four hour period for transmitting data back to Earth (downlink) with the remaining two hours for uploading the next set of observing commands (uplink). 
While the observatory and instruments were operational, the data processing pipeline needed to quickly process the eighteen hours of data to ensure the instruments and observatory were performing as expected. Time was critical in this phase as any issues found in the data could result in changes to either instrument or satellite operations.  
Failure to quickly process the observation data could result in lost observing days for the flight hardware and lost time was very expensive as the Herschel mission had a limited lifetime. The resulting data products generated in this phase of the mission were added to the HSA and were available to the instrument teams and observers within twenty-four hours of downlinking the data from Herschel.

An on-demand processing option was provided by the HSA allowing public users to reprocess existing observations using the HSA computing infrastructure. This processing option allowed users to reprocess observations with a new version of the data processing software than what was used when the observation was originally processed. Bulk reprocessing of the entire Herschel archive was done on a very limited basis.

Across all mission phases there was a need for sophisticated interactive analysis (IA) tools to analyze the test and observational data in order to perform deep dives into observatory or instrument issues. These tools provided a means for instrument calibration scientists to improve the quality the data products produced by the pipeline through data analysis and improved calibration products. This software was provided to the astronomers so they could use it to analyze their data.
The software was encapsulated in a tool called the Herschel Interactive Processing Environment (HIPE) \citep{sott2010}, see section \ref{sec:processing}.

\subsubsection{Standard Product Generation}
\label{sec:SPG}

The Standard Product Generation (SPG) framework provided a common basis for the instrument specific pipelines to be executed by the Herschel Science Centre.
This framework facilitated the aggregation of all the raw data taken by the 
observatory and any one of its science instruments 
to generate processed and calibrated data products that were free of instrumentation affects. 

\begin{figure*}[!ht]
   \begin{center}
   \includegraphics[height=6cm]{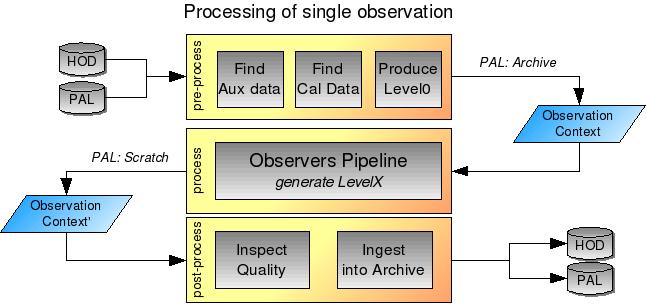}
   \end{center}
   \caption{Single observation data processing schematic}
   \label{fig:phases}
\end{figure*}

Figure \ref{fig:phases} shows the processing flow of Herschel science data.  
The SPG framework was composed of three main parts, a pre-processing (data aggregation) phase, the main data processing phase (observers pipeline), and a post-processing phase (quality and archiving).  
The raw instrument data was stored in an object database, the Herschel Object Database (HOD)\footnote{\href{https://www.cosmos.esa.int/web/herschel/legacy-documentation-observatory-level-3}{https://www.cosmos.esa.int/web/herschel/legacy-documentation-observatory-level-3}}, with the supporting instrument and observatory derived data being stored in FITS files accessible through the Product Access Layer (PAL) software \citep{li2012}.

A static initialization strategy was used to instantiate a set of plugins for the pre and post processing phases of the data processing pipeline. This allowed each science instrument to have their own strategy for collecting the required data for their pipeline and determining the quality and archival methods for that data. This decoupled and encapsulated the implementation details for these pre and post phases from the data processing logic to ensure a robust and consistent data processing environment. The benefit of the pre-processing step was that it provided observers with a self contained observation that had yet to be processed.  By downloading this pre-processed data product from the archive, observers if desired, could adapt the data processing algorithms (pipeline tasks) to meet their science goals in an interactive manner. This flexibility also gave observers choices in how they accessed their data (locally vs remotely). 

\subsubsection{Task package} 
\label{sec:taskDescription}

The data processing Task software package provided a common interface between the observatory and the instruments for encapsulating the instruments' data processing algorithms. The Task package was a core part of the software infrastructure provided to the instrument teams in order to build the software algorithms needed for data reduction. The Task package was designed to allow the instruments algorithms to be executed in all modes of operation such as quick-look analysis, interactive analysis or batch-pipeline processing.  


The Task\footnote{\href{http://herschel.esac.esa.int/hcss-doc-15.0/load/hcss_drm/ia/task/doc/index.html}{http://herschel.esac.esa.int/hcss-doc-15.0/load/hcss\_drm/ia/task/doc/index.html}} class encapsulated an algorithm and defined a standard signature for the inputs and outputs of that algorithm. With more than one thousand classes extending from the Task class, it was the most used data structure in the HCSS. The execution model of the Task was invoked with the perform method which effectively called the preamble, execute, and postamble methods of the Task. The execution model was fully configurable, i.e. each of the three methods were delegated via a strategy pattern \citep{gamma1994} and were replaced according to the different needs of the different environments. This was done at runtime and was completely transparent to both developers and users\footnote{\href{http://herschel.esac.esa.int/hcss-doc-15.0/load/hcss_drm/ia/doc/add/add.pdf}{http://herschel.esac.esa.int/hcss-doc-15.0/load/hcss\_drm/ia/doc/add/add.pdf}}.

\subsubsection{Herschel Interactive Processing Environment (HIPE)}
\label{sec:processing}

HIPE\footnote{\href{https://www.cosmos.esa.int/web/herschel/hipe-download}{https://www.cosmos.esa.int/web/herschel/hipe-download}} \citep{sott2010} is a publicly available open source, multi-platform stand-alone program providing access to all Herschel data processing tools enabling users to reduce and analyze observational data from all three Herschel science instruments. HIPE incorporated many processing toolboxes for analyzing/manipulating spectra (spectrum toolbox, spectrum fitter, and spectrum explorer), images (display toolbox) and spectral cubes (cube analysis toolbox) among many other processing tools needed by calibration scientists and useful to the general astronomical community. This program provided an interactive environment for developing, testing, debugging pipeline Task algorithms.

The HIPE program provided a powerful scripting editor and command line console allowing for script execution within HIPE.  HIPE also provided user access to all data processing functionality via a consistent user interface. Interactions with the graphical elements of HIPE were automatically translated and exported to the command line console in order that scripts could be generated and workflows automated.

\subsection{Software Implementation Languages}
\label{sec:softwareLanguages}

The Herschel mission selected the open source packages of Java\textsuperscript{TM} and Jython as the common software implementation languages for all data processing software. The benefits of this were to reduce cost, to ensure that the software could easily be reused across the instrument teams and that all developers were developing in a common language.
In addition, scripting support was provided in form of Jython - a Java\textsuperscript{TM} based interpreter of Python \footnote{\href{http://www.jython.org/}{Jython Home Page}} syntax. The scripting environment gave the benefit of using an interpreted language for rapid implementation and testing in the lab environment during instrument pre-flight validation as well as interactive analysis. The synergies between Java\textsuperscript{TM} and Jython allowed for easy access to methods implemented in Java\textsuperscript{TM},  when used in the interactive Jython console in HIPE. This gave non-developers (astronomers) an easier-to-use interpreted software language in which to manipulate the observational data and the more robust Java\textsuperscript{TM} based tools were developed by the software engineers. By the use of Java\textsuperscript{TM}, the developers were freed from the common software development problems associated with FORTRAN, C and C++, namely the memory management pitfalls and operating system dependencies. 

The selection of Java\textsuperscript{TM} and Jython in the late 1990s was a departure from the astronomy community's accepted standards of Interactive Data Language (IDL\textregistered) \footnote{\href{http://www.harrisgeospatial.com/SoftwareandTechnology/IDL.aspx}{IDL Home Page}} for image processing and CLASS \footnote{\href{http://www.iram.fr/IRAMFR/GILDAS}{GILDAS}} for heterodyne spectral processing. This resulted in additional overhead with regards to developing and validating the data processing algorithms as well as documenting and instructing users in how to use the provided functionality.

\subsection{Configuration Control}
\label{sec:configControl}
In the broadly distributed development team, it was imperative to have tight configuration control over the software and to test that any new software fit within the existing build. This was accomplished via the Herschel Continuous Integration Build (CIB) system. The CIB was an in-house cloud based build tool that would monitor commits to the Concurrent Versions System (CVS). When new commits were made, the system would checkout the new code, build it, and then execute all the dependent unit tests. If during compilation or execution of the unit tests, a failure was detected the build would be marked as failed and the module containing the new code would be marked as quarantined. The module would remain quarantined until a new commit was made to that module which did not fail in the build system.

Unit tests were written as part of the Herschel source code development process. The open source JUnit framework was selected to support the development and execution of all the Java\textsuperscript{TM} unit tests. Jython unit testing was supported through UnitTest but it was only used in a limited number of cases. The unit test code coverage goal of all Herschel Java\textsuperscript{TM} based software modules was eighty percent of the source code base. This goal was not enforced by the system and could not be achieved in all the modules - particularly, in modules that had been developed in the early years of the project before test coverage goals were formulated. 

In addition to unit testing, significant effort was placed on system testing. The instruments' pipeline was run on a daily basis for a preselected set of observations, using the most recent development stack maintained by the CIB system. The system tests did not prevent bugs from entering the software system, but they helped to identify when a side effect had been introduced into the system. All the artifacts, including datasets, logs and plots, generated by each test run of the pipeline were retained for later reference and debugging purposes. This provided a historical record of how the instrument pipelines were changing over time with daily granularity.

Astronomer acceptance testing was performed on development builds before these were released to the Herschel user community. The data processing and interactive tools were tested to ensure they supported a set of predefined work flows and that the results of these work flows were correct. As the observatory and instruments software teams did not have dedicated software quality assurance (QA) staff, this testing was done on a best effort basis. Much of this effort could have and should have been automated.

\subsection{Common Data Structures}

A common set of data structures for all instruments was required in order to facilitate data persistence and the transfer of data between processing sites and archives. These structures ranged from very simple wrappers around satellite telemetry to more complex structures such as spectra, images and spectral cubes.  All instrument software teams were allowed and more importantly, encouraged to extend these data structures enabling more instrument specific functionality. This made it easier to develop a common set of pipeline and analysis tools that manipulated observatory and instrument data in a standard way for all users. In the following sections, we will give a short description of the more important common data structures used by HIFI and their extensions used in data processing.


\subsubsection{Dataset}
\label{sec:dataStructure}

The Dataset was the primary data structure for bulk data within the Herschel software system and was specified as a Java\textsuperscript{TM} interface. The Dataset was a simple table with rows and columns, where a column was a n-dimensional array of numbers in which the lowest dimension determined the number of rows in the Dataset. In addition, the Dataset contained a map with key-value pairs of meta data. These items included elements such as the name of the observer, Right Ascension (RA), Declination (DEC), observation number, or other more specific elements needed by the data processing pipeline like the number of legs/steps in a raster map. 

The most common structures that implemented the Dataset interface were,
\begin{itemize}
\item{AbstractSpectrumDataset}

The AbstractSpectrumDataset was an extension of TableDataset - a concrete implementation of Dataset that allowed only columns of the same length. The AbstractSpectrumDataset contained at least two columns: a flux column that contained some measure of the intensity of the spectrum, and a column that contained the frequency or wavelength. Some optional but more common columns were the weight column, containing weights on the fluxes and the flag column which contained bit patterns (integers) that indicated various conditions applicable to the data points, see section \ref{sec:flags} on flags. 

An AbstractSpectrumDataset would contain one or more spectral segments, which provided a view on sub-sections of the spectra (spectral segments) that allowed for easy access and processing.

\item{Spectrum1d}

A Spectrum1d was an AbstractSpectrumDataset where the flux column was a 1-d column. It was the most simple variant of AbstractSpectrumDataset. In HIFI data processing, the Spectrum1d data structure was not used except for providing an exportable data structure for non Herschel data analysis tools. 

\item{Spectrum2d}

A Spectrum2d was an AbstractSpectrumDataset where the flux column had a 2-d structure. It was represented as a set of measurements of a 1-d spectrum. This structure was used extensively by HIFI, see section \ref{sec:htp}.

\end{itemize}

\subsubsection{Product}
\label{sec:product}

A Product was a container that contained references to zero or more Datasets plus history information and several required meta data fields. The Product data structure implemented a lazy loading mechanism for handling datasets, which was an important feature for a number of Herschel tools (e.g. plotting). The Product data structure became the primary structure for the serialization of Herschel data into FITS files \cite{sott2010} .  The Product meta data were stored in the header of the FITS file. A Product supported the concept of History, for auditing and data quality, in order to keep track of the algorithms (tasks) acting on the data contained within the Product.  


A Context was a special type of Product that could hold references to other Products and Contexts. This allowed for the construction of arbitrary complex nested Product structures. The MapContext and ListContext directly extended the Context class.

The following are a list of some of the more widely used Herschel Products:

\begin{itemize}
\item{ObservationContext}

The ObservationContext was the main container that encapsulated all the data contained in a single Herschel observation. It contained all the measured and derived data needed by the data pipelines to process an observation as well as the output of the data processing pipeline. For HIFI the data processing outputs were organized as the level 0 (section \ref{sec:level0}), level 0.5  (section \ref{sec:level0.5}), level 1 and 2  (section \ref{sec:level1}) and level 2.5  (section \ref{sec:level2.5}) products.

\item{AuxiliaryContext} 
\label{sec:auxtree}

The Herschel auxiliary data tree was a data structure that contained all the spacecraft specific data needed by the  pipelines to process an observation. The observatory pointing Product and the Spacecraft/Instrument Alignment Matrices (SIAM) were examples of auxiliary products that were needed by the pipeline to properly associate the motion of the observatory and instrument with the generated observational data. Further details regarding the contents of the Herschel auxiliary products can be found online in the Herschel Explanatory Legacy Library\footnote{\href{https://www.cosmos.esa.int/web/herschel/llegacy-documentation-observatory}{https://www.cosmos.esa.int/web/herschel/legacy-documentation-observatory}\label{hellobs_fn}}.

\item{SimpleImage } 

The SimpleImage was the container used to collect the final result for mapping observations without a spectral resolution. It contained a World Coordinate System (WCS) and one or more two dimensional arrays that represented typically observed or processed fluxes, wave, weights and flags of the region of the sky observed.

\item{SpectralSimpleCube } 
\label{sec:simplecube}

The SpectralSimpleCube extended SimpleImage and was the container used to collect the final results for spectral map observations. It contained a WCS and one or more three dimensional arrays that represented typically observed or processed fluxes, wave, weights and flags. The first two axes represented the RA and DEC coordinates along the observed sky positions, while the third axis represented a spectral axis. 


\item{SimpleSpectrum }
\label{sec:simplespectrum}

This was the  container used to store the final HIFI results for point mode observations. SimpleSpectrum was Product that encapsulated a Spectrum1d dataset.  It was a data structure more commonly used outside of the Herschel software environment and was used to make it easier to use Herschel data with non-Herschel tools.
\end{itemize}

\subsection{Common Software Libraries}
\label{sec:infra}

The HCSS included a large set of common software tools which were required by the instrument teams to process their data.  These tools evolved over time as the phase of the mission changed and were adapted by each instrument to meet their needs. The calibration and science data generated and returned by each instrument was unique to that instrument. However, whenever possible common software elements were shared between each instrument and the observatory so as to not duplicate effort such as,    

\begin{itemize}
\item {\bf Numerics}: An arithmetic toolbox which provided N dimensional array definitions, functions and procedures to operate on these arrays 
\item {\bf Product access layer}:  Tools to read/write products locally from disk or a database and interact with HSA
\item {\bf Plotting libraries}:  Tools and plotting functionality to display column data.
\end{itemize}

The spectrum flagging and arithmetic libraries were used extensively by the HIFI pipeline and are described below in more detail.

\subsubsection{Data flagging/masking}
\label{sec:flags}

Each data point within an observation had a data flag associated with it. These flags were 32-bit integers, where each bit indicated the presence or absence of some predefined condition. These flags could indicate bad pixels, saturation or the possible presence of a spur as an example. Some flags were considered severe and as such the associated data points were excluded in any subsequent calculations, other flags were merely warnings for the user. Additionally, row flags were associated with conditions required for a valid complete measurement and again some row flags were considered severe and others were only a warning for the users. 

\subsubsection{Arithmetic operations on spectra}

Spectrum arithmetics toolbox was at the core of HIFI pipeline, see section \ref{sec:software}. The spectral arithmetic operations (addition, subtraction, multiplication, division) were applied on complete spectral components. The spectrum arithmetics toolbox was a library that was not only provided for arithmetic manipulations on the flux values of spectra but could also be configured to incorporate consistent operations on weights (noise) and flags and adjustments of suitable meta data. For example, flags were propagated to the resulting spectra by preserving the information specified per data point.

The spectral arithmetics operations were designed to work with Herschel spectra, in particular, with any data structures that implemented the SpectrumContainer interface such as Spectrum2d.   Accordingly, the spectrum arithmetics toolbox operated on data of all three instruments. 

The following operations were provided by the toolbox:

\begin{itemize}
\item \textbf{Basic arithmetic operations}: Add, Subtract, Multiply, Divide. The basic arithmetic operators (\mysingleq{+}, \mysingleq{-}, \mysingleq{*}, \mysingleq{/}) were overloaded in Jython. When writing Jython scripts or when implementing tasks in Jython, two spectrum containers (spectra1, spectra2) could be added simply by writing
\begin{equation}
result = spectra1 + spectra2
\end{equation}
The result contained the point-wise added point spectra found in both of the input containers. For exploiting more advanced configuration possibilities, the instances of the underlying classes needed to be used. As an example, the operations could be restricted to a subset of the spectra.

\item \textbf{Spectrum manipulation tools}: Select, Extract, Replace, Stitch. The {{\it select}} allowed to efficiently filter spectrum containers based on any characteristics defined for the point spectra. With the {{\it  extract}}, the spectra were cut to a suitable size (defined in the wavescale - e.g. frequency range or number of frequency bins). The {{\it replace}} allowed for a combined cut and paste and the {\it stitch} provided a powerful tool to combine spectra possibly overlapping or even defined at different frequency scales. 

\item \textbf{Further operations}: Average, Smooth, Resample, ConvertWavescale. The {\it average} computed a simple arithmetic mean or the weighted average of multiple spectra per frequency bin. With the {\it smooth}, the spectra could be smoothed along the frequency axis with several smoothing kernels. {\it Resample} allowed the spectra to be resampled to any not necessarily uniformly spaced frequency grid. In the HIFI pipeline, this was used extensively before steps that combined spectra. The {\it convertWavescale} was used to transform the data between different physical units of the wave scale (frequency, wavelength, wavenumber).

\end{itemize}

\section{HIFI Instrument Data} 
\label{sec:structures}

The HIFI instrument consisted of two independent spectrometers, the acusto-optical Wide Band Spectrometer (WBS), and the High Resolution Spectrometer (HRS) autocorrelator. The data generated by these spectrometers were in the form of data packets called Telemetry. In the pre-flight phase of the mission the data generated by these spectrometers were streamed directly to the database in the laboratory, however during the operational phase of the mission, the data packets were transmitted to the ground station and stored as plain text files.  The Telemetry packets came in two varieties, housekeeping (HK) packets and science packets. There are a series of hardware to software interface documents describing the structure of these data packets with the most important ones being, \citep{HPACKET} and \citep{HGROUND}. For the HIFI instrument the data packet structure is described in \citep{HIFIPACKET}. 

All packets, science and HK, contained a building block id (BBid), a unique number which indicated the purpose of the packet. The BBid numbers were used by the pipeline to determine how a particular set of data should be processed in the pipeline. 

HK and science data were generated on-board the observatory and instrument and were sent to the ground station by two asynchronous processes. To enable the association of the HK and science packets for data processing, a unique counter was introduced into the data frame packets to mark the relevant HK packets. HK packets contained information about the health of the observatory and the instruments as well as information relevant for interpreting the science data.

The science data packets were assembled into Dataframes in the pre-processing phase of the pipeline, see section \ref{sec:level0}. The Dataframes represented the channel readouts for one measurement of either the WBS  or HRS spectrometer and one for each polarization. The science packets were required to be transmitted by the instrument and received by the ground station in sorted order. Each of the science packets contained a sequence number to check for missing packets. Missing and/or corrupted packets occurred in some limited occasions and the data processing pipelines identified and flagged these Dataframes as bad so that they were ignored in the subsequent processing steps, see section \ref{sec:quality}. Dataframes with the  same BBid were collated into a single HifiSpectrumDataset.

\subsection{HIFI Data Structures}
\label{sec:htp}

The HIFI data structures were extended from the data structures defined in section \ref{sec:dataStructure}. 
The use of common HCSS data structures made it easier to process HIFI data in a standard way and enabled HIFI users to use the Herschel developed software tools with their data. The following data structures were the most common structures used by the HIFI pipeline.

\begin{itemize}
\item{HifiSpectrumDataset}

A HifiSpectrumDataset (HSD) was extended from the Spectrum2d and specialized for the HIFI instrument. Each row in the HSD contained information about one HIFI Dataframe for an observation. The HSD contained a large number of additional data columns needed by HIFI pipeline. 
A HSD was constructed for all subsequent measurements of the HIFI WBS/HRS that shared the same BBid. The BBids were used to distinguish parts of an observation which had the same function (e.g. wavelength calibration, dark current measurement, on source, off source, etc.).  
A single HIFI observation could contain thousands of HifiSpectrumDatasets depending on the type and length of the observation.

\item{HifiTimelineProduct}
\label{sec:htp}

A HifiTimelineProduct (HTP) was an extension of the Product class and was designed to contain zero or more HifiSpectrumDatasets and a SummaryTable. The datasets were grouped into \textit{boxes} of datasets within the HTP adding another layer of abstraction. When the HTP was originally designed and implemented each HSD was wrapped as a single Product and therefore a single FITS file. To avoid the proliferation of a large number of small FITS files with an according drawback on I/O performance, groups of a configurable number of HSD's (by default 100) were packed into \textit{boxes} - each \textit{box} was stored in a single FITS file.

\item{SummaryTable}

A SummaryTable was a small table contained within the HTP that summarized the contents of an HTP, see figure \ref{fig:summarytable}. There was one row in this table for every HSD contained in the HTP. This table contained some key items needed for a quick survey of the observation, namely the \textit{type} of HSD along with its identifying \textit{Bbid} number (this column really should have been called bbType as the Bbid is actually a combination of the building block type, which is shown in the column, and building block execution order number or bbNumber). The column \textit{isLine} was used to indicate a science data block and whether it was on-source (\textit{true}) or on reference (\textit{false}). Other columns included \textit{isHrs} (data from the HRS), \textit{isWbs} (data from the WBS), the \textit{fullName} of the instrument command that created the data, the \textit{start} position in the sequence of Dataframes and the \textit{length} of Dataframes in that sequence.  

Similar SummaryTables can be found alongside the PACS and SPIRE main data products.

\begin{figure}[!ht]
   \includegraphics[width=\linewidth]{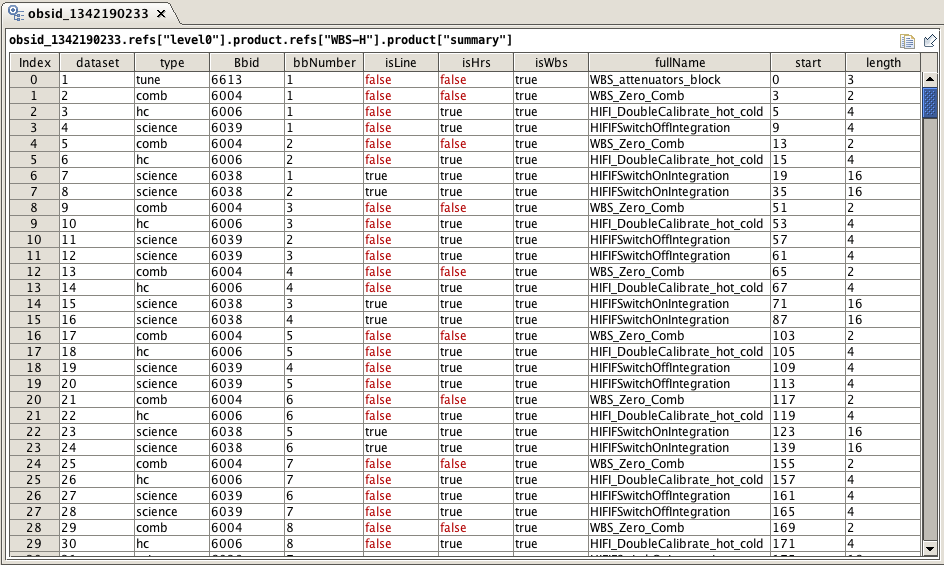}
   \caption{Level 0 HifiTimelineProduct summary table for the WBS horizontal polarization}
   \label{fig:summarytable}
\end{figure}

\end{itemize}

\subsection{HIFI Calibration Data} 
\label{sec:caltree}

The HIFI instrument calibration was contained in the HIFI calibration tree.  In this product resided all the measured and derived instrument specific data needed by the HIFI pipeline to process an observation. 
At the top level, the calibration tree was a map that could contain both additional maps (calibration tree nodes) or a concrete calibration object. This structure allowed calibration objects to be organized in logical groups associated with their use in the HIFI pipeline. A calibration object was a Product that contained one or more tables and associated meta data about those tables, see section \ref{sec:dataStructure} on data structures. 

The calibration tree was equipped with a versioning mechanism and the pipeline could be configured to access specific versions. In the course of the mission, several versions were released by the calibration scientists which incorporated the most recent knowledge about the instrument. This mechanism provided full traceability and reproducibility by providing the version of both the software and of the calibration tree which was of particular importance when publishing results.

\begin{figure}[!ht]
  \begin{center}
   \includegraphics[width=\linewidth]{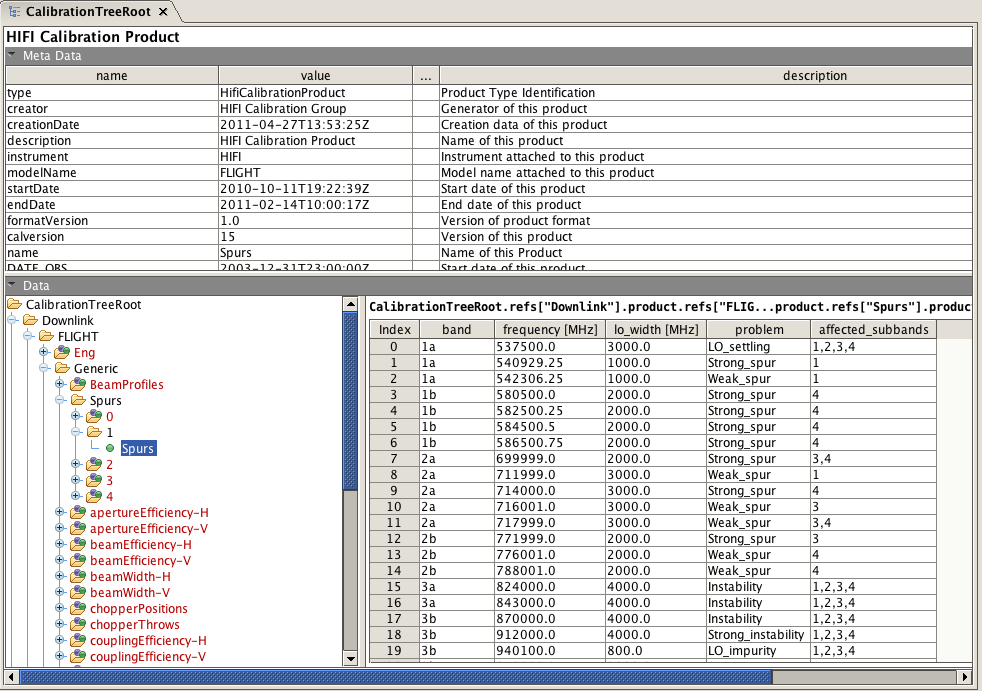}
   \end{center}
   \caption{HIFI Calibration Tree, Spurs Calibration Product - Version 15. This image is divided into three sections with the top section showing the meta data associated with calibration product, the lower left section showing a hierarchical view of the calibration tree and the lower right section showing a table containing data about a particular calibration product, in this case, a Spurs table identifying the frequency, width and type of spurs found.}
   \label{fig:state}
\end{figure}

During the pre-processing stage of the HIFI pipeline, see section \ref{sec:SPG}, the observation context was populated with the configured version of the calibration tree (by default the latest version). 
This was accomplished by a dedicated {\it plugin} (CalPlugin) that directly plugged into the SPG framework. The CalPlugin would not only select the correct version of the calibration tree to be attached to an observation, it would also modify the structure of the tree depending on the observation.  The observation-associated-calibration tree only contained information relevant to that observation.
This tailoring of the calibration data limited the total observation data size, hence reducing the download time of an observation from the archive. This type of customization was unique to the design of the HIFI calibration tree.  A view of the HIFI calibration tree root node can be seen in figure \ref{fig:state}. 

For each production release version of the calibration tree there was a matching developer release. The developer release was a type of staging area for testing new versions of calibration objects. This allowed for easy testing and modification of the objects in the tree. The history of the all the changes made to the developer branch of the tree was not retained, only changes published to the production branch were persisted. Once a developer release was deemed ready for production it was published, receiving a version number in the master root node and could no longer be modified. 

The calibration tree framework consisted of a set of basic generic instrument independent objects and operations. The following section is a description of those objects extended and used by HIFI. 

\subsubsection{HIFI Calibration Tree Object Types}
The HIFI calibration tree had three basic data structures which allow it to store any type of HIFI instrument calibration data. 
The first was the basic HifiCalibrationProduct.  
The basic calibration product implemented the HistoryIdentifiable interface which allowed the pipeline to identify that a given calibration product had been applied to the data.
Next was the HifiListContext
 which represented a time series of calibration objects. This type of object was important when different calibration values needed to be applied to observational data depending on when the observation was taken (early vs late mission). 
Finally, the HifiMapContext 
 represented a HashMap of calibration objects. This type of calibration object was useful when one needed to specify a particular correction for a single observation or group of observations. These types of corrections were typically applied to observations where a hardware malfunction occurred or in the case of the electrical standing wave corrections \citep{2014AIPC.1636...62K}, as each correction was unique to the observation which it applied.

When updating or adding a new calibration object to the HIFI calibration tree there were contractual obligations to fulfill before the system would allow the changes to be applied to the developer branch of the tree. This was to protect the integrity of the tree ensuring consistency across updates and to allow for proper logging and versioning when a particular object was applied to the observational data. Each calibration object had a set of mandatory meta data that had to be specified in order to properly identify it within the system. A set of predicate rules were applied to ensure that only correctly formatted calibration objects were added to the tree. These included items like version number, name, description, applicable start and end dates. The calibration framework had methods that allowed individual products to be fetched using this information. This was particularly helpful for the pipeline tasks ensuring that it could quickly and easily find the correct object to apply to the observation data during processing.  This also allowed the pre-processing CalPlugin to assemble a unique calibration tree for a given observation ensuring a light weight structure.

\subsubsection{Calibration objects in an HIFI ObservationContext}
\label{sec:sec:calObs}
\begin{figure}[!ht]
   \includegraphics[width=\linewidth]{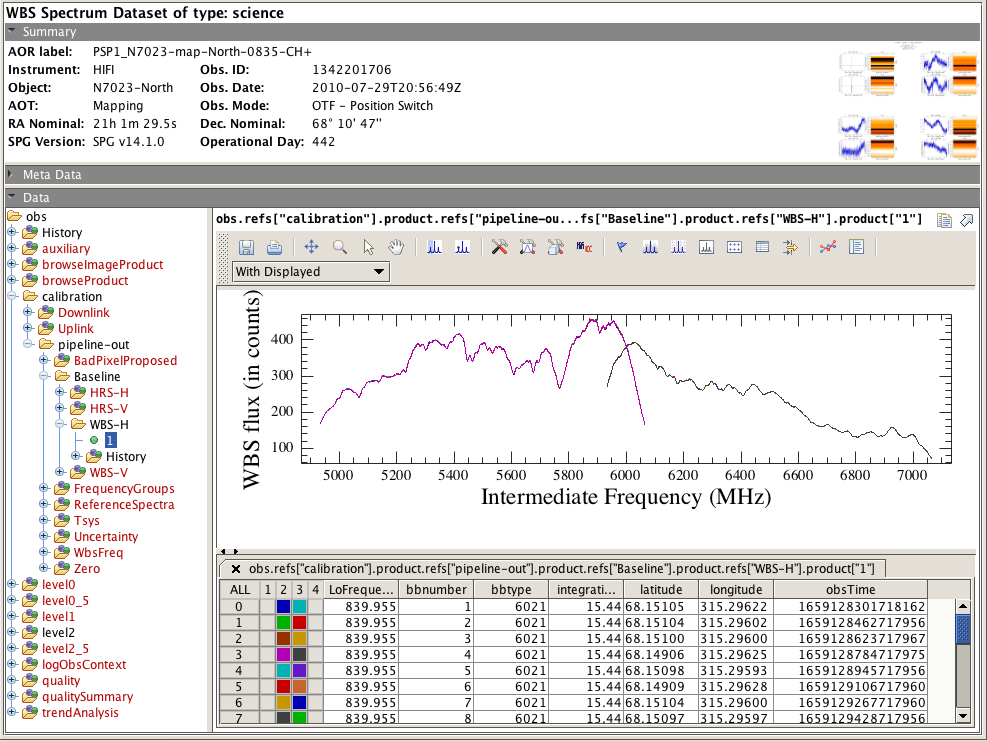}
   \caption{HIFI observation with the calibration tree as seen in HIPE.  The lower left panel, \textit{Data}, shows the observation with the calibration tree.  The lower right panel, with plot, represents the data highlighted in the \textit{Data} panel as viewed in the spectrum viewer.  The spectra shown can be (de-)selected via the lower right sub-panel.}
   \label{fig:calobs}
\end{figure}

Figure \ref{fig:calobs} shows a representation of the HIFI calibration tree connected with an observation as seen within HIPE.   The top panel provides general information about the observation (name, observed date, position, etc.).  The lower left-hand panel shows the observation context tabs with the calibration tab opened.  The lower right panel shows the calibration data: for this case the pipeline-out/Baseline/WBS-H calibration data.  The calibration attached to each observation contained three main branches, downlink, uplink and pipeline-out. 
\begin{itemize}
\item The downlink branch represented calibration data that were applied to the science data.  The downlink calibration data had been independently measured or derived from several calibration observations and were not necessarily based on any one observation, e.g. load coupling coefficients or sideband ratio \citep {shipman2017}.  
\item The uplink branch contains calibration data that was used to tell the HIFI instrument how to perform a particular observation, e.g the amount of time to observe at a particular location or the expected signal noise from the resulting observation \citep{HIFIH}.  Many of these parameters were determined at the planning stage of the original observation request. 
\item The pipeline-out branch contains calibration data that had been generated by the  pipeline during observation processing, e.g the system noise temperature (\Tsys). These data were useful in determining other statistics about the observation as part of further data analysis. 
\end{itemize}

\subsection{Trend Analysis Products}
\label{sec:trend}

In order to assess the behaviour of the instrument over time, parameters were collected within every observation and combined with data from other observations to form a set of trend data over the entire mission.  
The trend analysis data containers were used by HIFI calibration scientists to check various aspects of the observation and the data processing results. 
Some of the data objects contained in the trend section could be combined with other similar data objects from other HIFI observations to construct trend plots to monitor instrument performance over time. 
This was done with the data contained in the T\textsubscript{sys} data container, as an example. 

Each pipeline processing Task could generate and save data to the TrendAnalysisContext.  
A TrendAnalysis data object was simply a generic MapContext that could contain any type of Herschel data product.   
The data containers present in the HIFI TrendAnalysisContext are briefly described below:

\begin{itemize}
\item{ Frequency trend of WBS}

The frequency response of the CCDs  in the  WBS instrument was computed during each observation through periodic  artificial spectra (COMB) with a series of stable frequencies at 100 MHz steps.  Detailed information of each COMB line and their fit results were stored inside the Quality\_level0\_5 product, useful for calibration scientists in order to determine if an error had occurred during a specific fit in the data reduction of an observation.  

\item{FPU housekeeping trend}
  
The FpuTrendProduct  contained tables with HK parameters from the Focal Plane Unit (FPU) as a function of time for an observation. These parameters and some of their combinations were monitored against specified thresholds and when those constraints were violated quality flags were raised.
The DoHkCheckTask allowed calibration scientists to include additional HK parameters for monitoring without requiring pipeline software changes. Utility methods were included to help the calibration scientists choose which HK parameters to monitor and set which quality flags needed be raised when out-of-limit conditions occurred.

\item {LOU housekeeping trend}

The  LoTrendProduct contained  HK parameters from the Local Oscillator Unit (LOU).  Both the FpuTrendProduct and the LoTrendProduct content were used in the level 0 pipeline in order to flag possible out-of-limits on specific HK parameters. When this occurred, a dedicated quality flag would be raised and added to the quality product.

\item{TMpageContext}

The TMpageContext contained diagnostic tables for every local oscillator (LO) tuning that was performed during the observation. This information was mostly of interest for instrument scientists, and offered a selection of LOU HK parameters during predefined steps of the tuning process, at time intervals finer than the sampling rate offered for the periodic HK compiled in the FpuTrendProduct or the LoTrendProduct contexts.

\item{Spur table  (WBS only)}

The SpurTable was a table of spurious signals detected in the cold load spectra used for the intensity calibration of the spectra. A spurious signal could be created by an impure local oscillator signal and would appear as a data spike over several channels.  

At the end of the level 0.5 pipeline, spurs were identified when the signal in raw counts was above the saturation level of 800 counts (pre-bandpass calibration) and appeared in the data as a Gaussian-like feature.  A region 1.5 times broader than the width of the spur was flagged. 
This saturation level could be adjusted by the user. Depending on the saturation level, the channels were flagged indicating that data were corrupted in a given range and that the algorithm could not determine a good fit to a Gaussian profile.

 \item{StatisticsContext}

The StatisticsContext contained a series of tables with computation of the first momentum of the observed spectra (mean and standard deviation) as well as the median. It was provided for each spectrometer, in each subband of the given spectrometer, and for the datasets at level 1 and level 2. Only saturated pixels are excluded from the computation.  

The StatisticsContext was used to identify whether the observation resulted in the predicted signal to noise as given by HSPOT during the planning of the observation \citep{shipman2017}.

\item{T\textsubscript{sys}  Context}

The T\textsubscript{sys}Context contained the T\textsubscript{sys}TrendTable per backend and per subband, and provided the LO frequency, the central intermediate frequency (IF), the observation time, (nominal) resolution for the backend, the double-side-band system noise temperature and associated standard deviation computed from the hot/cold load datasets.
\end{itemize}

\section{HIFI Data Processing Software}
\label{sec:software}
  
The HIFI data processing software was built using the SPG and Task frameworks, see sections \ref{sec:SPG} and \ref{sec:taskDescription}. The overall HIFI pipeline was designed following a top down approach with the goal of being able to process all HIFI instrument data using a simple interface for all users. Along with providing a simple interface to users, the pipeline needed to be highly customizable in order to support a multitude of use cases while allowing for change as the HIFI instrument became better understood.

The HIFI pipeline could process any HIFI observation with a single command and one or more configuration parameters supplied by the user to the Jython interpreter.  The simplest form of this command required only the observation identifier (obsid) parameter,  \textit{obs = hifiPipeline(obsid=12345678)}. When executing this command, the observation context was fetched from HSA and the execution of the pipeline was performed locally. Additionally, similar commands could be applied to process any pre-flight observations, calibration observations or manually commanded observations.

The HIFI pipeline processed an observation in a series of steps (represented as tasks) and those steps were grouped into a series of levels (see Appendix A of \citep{shipman2017} for flow diagrams).
%
%
The HIFI pipeline contained the following processing levels, 
\begin{itemize}
\item \textbf{Level 0}: The pre-processing phase of the pipeline where all the necessary inputs were collected.
\item \textbf{Level 0.5}: Removing instrumental artifacts associated with the WBS and HRS spectrometers.
\item \textbf{Levels 1 \& 2}:  Removing observational artifacts associated with the different observing modes and applying the calibration.
\item \textbf{Level 2.5}: Performing higher level processing tasks which were observing mode specific.
\end{itemize}

At the top level HifiPipelineTask invoked a series of sub-pipelines representing the data processing level and taking the raw observation data to any configured processing level. Each sub-pipeline at a given processing level was composed of many tasks, each one representing an algorithm that performed an important transformation on the observational data.  The sub-level pipelines were implemented as Jython tasks allowing them to be configurable by the users whereas the HifiPipelineTask and most of the tasks invoked by the sub-level pipelines were implemented in Java\textsuperscript{TM}.

\subsection{HIFI Pipeline: Configuration and user interaction}
\label{sec:config}
The main operation mode of the HIFI pipeline was to run in a standard \textit{hands-off}  fashion within the SPG framework where no interactions were possible.  The pipeline had an interactive mode for the users as well.  The main requirements of the users on the HIFI pipeline were,

\begin{itemize}
\item  Users must be able to change the order in which the sub-level pipeline steps were invoked without having to recompile the code and generate new builds. The different processing levels were considered fixed.
\item  Users must be able to replace or modify individual pipeline steps (tasks).
\item  Users must be able to customize the pipeline from the command line or a graphical user interface (GUI) and that the graphical approach should generate a fully functional script that could be directly substituted for any GUI interaction.
\end{itemize}

These requirements were achieved by structuring the HIFI Pipeline code in the following way, 

\begin{itemize}

\item The sub-level pipelines were implemented as Jython tasks and invoked a suitable sequence of data processing steps that represented the data flow within the pipeline. The customizability was achieved by allowing the user to replace the sub-level pipelines with a customized Jython script where the sequence of pipeline steps could be modified, single processing steps replaced, or parameters passed to the steps modified. The Jython interpreter could be used to parse and execute the pipeline script.

\item  Custom processing steps (tasks) could be provided in form of Jython or Java\textsuperscript{TM} tasks. Typically, they were implemented in Java\textsuperscript{TM}, however during development and for rapid prototyping or for prototyping by instrument scientists, Jython implementations of many of the tasks were created. Each Task had its own GUI where inputs to the Task could be modified.

\item The HifiPipelineTask provided a GUI that collected the input parameters of all of the lower level Task GUIs into a single panel.  The GUI for the pipeline did not allow for a full customization at all the pipeline levels - such as for changing the sequence of tasks, however it did support the modification of the parameters passed to the processing steps. The level 2.5 processing level supported the adding and removal of tasks as well as their reordering within the sub-level pipeline itself. When a Task was executed via the GUI, the Task returned an output of all the input parameters to the Task allowing users to generate fully functional scripts.  
\end{itemize}

\subsection{Level 0} 
\label{sec:level0}
The main goal of the level 0 pipeline was to prepare a consistent and complete HifiTimelineProduct - actually, one for each spectrometer and polarization - ready to be processed in the subsequent pipeline levels. This required a detailed knowledge
about how the data frames and the source packets were generated by the instrument, see section \ref{sec:structures}. In addition, satellite pointing information was added and some sanity checks were applied to flag any occurrences where instrument house keeping parameters fell outside of specifications.

This part of the HIFI pipeline was different from the other pipeline levels in that the possibility to customize this pipeline was limited. The level 0 pipeline could be re-processed for instance with different auxiliary or calibration data  (e.g.  pointing correction), however the Task parameters themselves could not be changed.

All the tasks included in the level 0 pipeline could be accessed within HIPE. The expert user could retrieve the HifiTimelineProduct from level 0, apply the desired tasks with the chosen parameters and then reinsert the HifiTimelineProduct back into ObservationContext in order to update level 0 data product, if needed.

\subsection{Level 0.5}
\label{sec:level0.5}

The second data processing level reached by the HIFI pipeline was called level 0.5. The component pipelines that made up this level were strictly related to the spectrometer data to be processed. There were two independent pipelines that formed the basis of the level 0.5 pipeline. A pipeline for the Wide Band Spectrometer (WBS) and a pipeline for the High Resolution Spectrometer (HRS). In reality there were actually four spectrometers when taking into account the horizontal and vertical polarizations of the signal, therefore after executing the level 0.5 pipeline four unique HifiTimelineProducts were created. 


These pipelines were designed to produce data products that were equivalent in data structure and calibration so that data from either spectrometer could be processed in a standard way when executing the level 1 pipeline. These pipelines removed most of the instrument specifics from the data to achieve data compatibility.  The HifiPipelineTask decided which spectrometer pipeline to use based on the meta data contained in the level 0 HifiTimelineProduct. Likewise, each polarization had its own calibration table and again the HIFI pipeline would decide which calibration table to use based on the meta data.

The tasks in WBS and HRS component pipelines had a fixed order for all observations. The user interaction with these pipelines when doing interactive analysis with HIPE was limited. In principle, user interaction was possible by adopting the same generic mechanisms described in section \ref{sec:level0}, however it was rarely applied. To save disk space and reduce download time, the HifiPipelineTask could be configured to remove the level 0.5 data after successfully completing the level 1 pipeline. By default, this option was activated.

\subsection{Levels 1 and 2}
\label{sec:level1}

The purpose of the level 1 pipeline was to flux calibrate the HIFI data using the internal load measurements and to subtract the background by combing observations from different positions on sky (on/off target, source/ref positions).  Additionally, the frequency scale was corrected for the motion of the spacecraft and was transformed into a frequency scale consistent with the Local Standard of Rest (LSR) reference frame (for fixed targets) or to the reference frame of the moving targets (for solar system objects). 

In the level 2 pipeline, the intensity calibration was finalized by applying the antenna temperature and sideband correction to bring the wave scale to the true physical frequencies. The resulting spectra were averaged where applicable - i.e. where the spectra refer to the same position (or object in case of solar system objects) on sky and the same frequency scale.

While in the level 0.5 pipeline the instrument specific artifacts were eliminated, in the level 1 and level 2 pipelines the observing mode specific details were considered. Eighteen different observing modes were designed to efficiently resolve the spectral signal \cite{shipman2017}. The observing modes correspond to different schemes for switching between the source position and suitable reference positions on sky or on the telescope or between different frequencies. These schemes lead to different patterns in the sequences of data frames collected during an observation. The complexity in the level 1 and level 2 pipelines was mostly related to dealing with reducing these patterns by combining them into intensity and frequency calibrated spectra.

\subsubsection{Managing the Complexity Introduced by the Variety of Observing Modes} 
\label{sec:pipelineConfig}

The HIFI instrument allowed observers to plan their observations against a set of astronomical observing mode templates (AOTs)\citep{shipman2017, ossenkopf2003}. These AOTs fell into three main categories, Single Point Mode, Spectral Mapping Mode and Spectral Scan Mode. Each of these categories contained several templates for an overall total of eighteen different AOTs or observing modes.

For each observing mode, specific pipeline processing logic was applied, therefore conceptually each observing mode came with its own pipeline. Instead of implementing a dedicated pipeline for each observing mode, the work flow logic of all observing modes was mapped to a single pipeline, with a few observing mode specific switches in the Jython scripts that defined the pipeline work flow. In addition, the tasks of the level 1 and 2 pipelines could be configured by loading AOT specific configuration objects containing a set of input parameters for each Task. This observing mode specific pipeline configuration was loaded as the first step of the level 1 and 2 pipeline. This configuration was found in either XML files shipped with code as the first step in the pipeline script (for automated batch processing) or from the GUI when running the pipeline in an interactive mode.

In principle, the behaviour of these pipelines could be changed by simply modifying a XML file - which was rarely done.  This design allowed for better code re-use and avoided code duplication. Specifically, we did not have to maintain separate pipelines for each observing mode and dedicated pipeline scripts. On the opposite side, the design did not easily support adding new observing modes with completely different work flows. Since the available observing modes were fixed and given by the design of the HIFI instrument, this software design was judged as reasonable and sufficiently flexible.

\subsubsection{Main Functional Blocks}

For a detailed description of the various data reduction steps applied in level 1 and 2 parts of the pipeline please refer to \citep{shipman2017}. We have just summarized the main functional blocks to point out a few additional details below,  

\begin{enumerate}[label=\bf(\alph*)]
\item {\bf Sanity Checks}:
At the beginning of the level 1 pipeline, several data sanity checks were conducted. The data had to be checked whether it complied with the expected structure of the observing mode such as whether housekeeping data assigned to the data frames (such as buffer, chopper position, LO frequency) followed the expected patterns (checkDataStructure, checkFreqGrid, checkPhases). In cases where inconsistencies were identified, dedicated quality flags were raised and added to the observation context.

\item {\bf Intensity Calibration at level 1}:
The information included in the data frames in form of flux counts were transformed to a physical intensity scale attributed to the signal from the observed object. All the spectra were divided by a bandpass that was constructed from data frames observed from hot/cold sources mounted on the telescope. This resulted in transforming the spectra from simple flux counts to a physical intensity (temperature) scale (mkFluxHotCold, doFluxHotCold). Depending on the observing mode, suitable subtraction schemes were applied to remove a dark sky reference signal. In this way, the part of the signal that could be attributed to the original source of interest would be isolated (doRefSubtract, mkOffSmooth and doOffSmooth). 

\item {\bf Velocity Correction}:
The frequency scale was adjusted for the motion of the spacecraft. For solar system objects, the frequencies were transformed to the rest frame of the object - in all other cases, the frequencies were transformed to the LSR frame. 

\item {\bf Intensity Calibration at level 2}:
Further processing steps were applied for finalizing the intensity calibration: a sideband gain correction was applied and the spectra were brought to the antenna temperature scale (mkSidebandGain, doSidebandGain, doAntennaTemp). 

\item {\bf Uniformly Gridded Sideband Frequency Scale}:
The frequency scale was converted to the sideband frequency scale and the frequencies were resampled to a uniform frequency grid (frequencyConverter, mkFrequencyGrid, doFrequencyGrid). 

\item {\bf Average}:
The spectra with a common underlying frequency range were averaged (doAvg).
\end{enumerate}

The application of all these pipeline steps resulted in level 1 spectra  (including the corrections described in [a-c]) and level 2 spectra ([d-f])  along with extra generated calibration, quality, trend and statistics products on the distribution of the spectra found in the HifiTimelineProduct. All data products generated were linked to the ObservationContext. In cases where inconsistencies were identified, dedicated quality flags were raised and added to the ObservationContext as well.

\subsection{Level 2.5}
\label{sec:level2.5}

The Level 2.5 pipeline was the final step in the HIFI data processing pipeline. This step processed the level 2 HifiTimelineProducts into their final state and like the level 1 and 2 pipelines depended on the observing mode of the observation. In the case of point mode observations, the final result from the pipeline was a single spectrum from the average of all the observed spectra. 
For mapping modes this was a spectral cube and for spectral survey modes this was a single deconvolved spectrum, see section \ref{sec:product}.  
All of these products required further astronomer interaction from this point on in order to produce higher quality scientific results and were not easily automated with further pipeline steps. 
The level 2.5 pipeline was constructed as a best effort for automated processing, however better results could be achieved. 
As an example observational data could be improved with baseline subtraction which required \textit{interactive help} from astronomers before being applied to the spectra. 
Whenever possible, expert-user generated interaction was automated in the pipeline such as flagging spurs and other bad data in spectral scans. This information was accordingly added to the calibration tree and applied by the pipeline during bulk reprocessing.  

This last level of data processing required more flexibility, configuring and ordering of the tasks in the pipeline. 
In addition, some of the tasks combined data products from different spectrometers and polarizations which was outside the scope of the standard pipeline steps.  
Finally, the key information about the observation was summarized including an estimate for the root mean square (RMS) of the data. 

\begin{figure*}[!ht]
  \begin{center}
   \includegraphics[width=16.5cm,height=13cm]{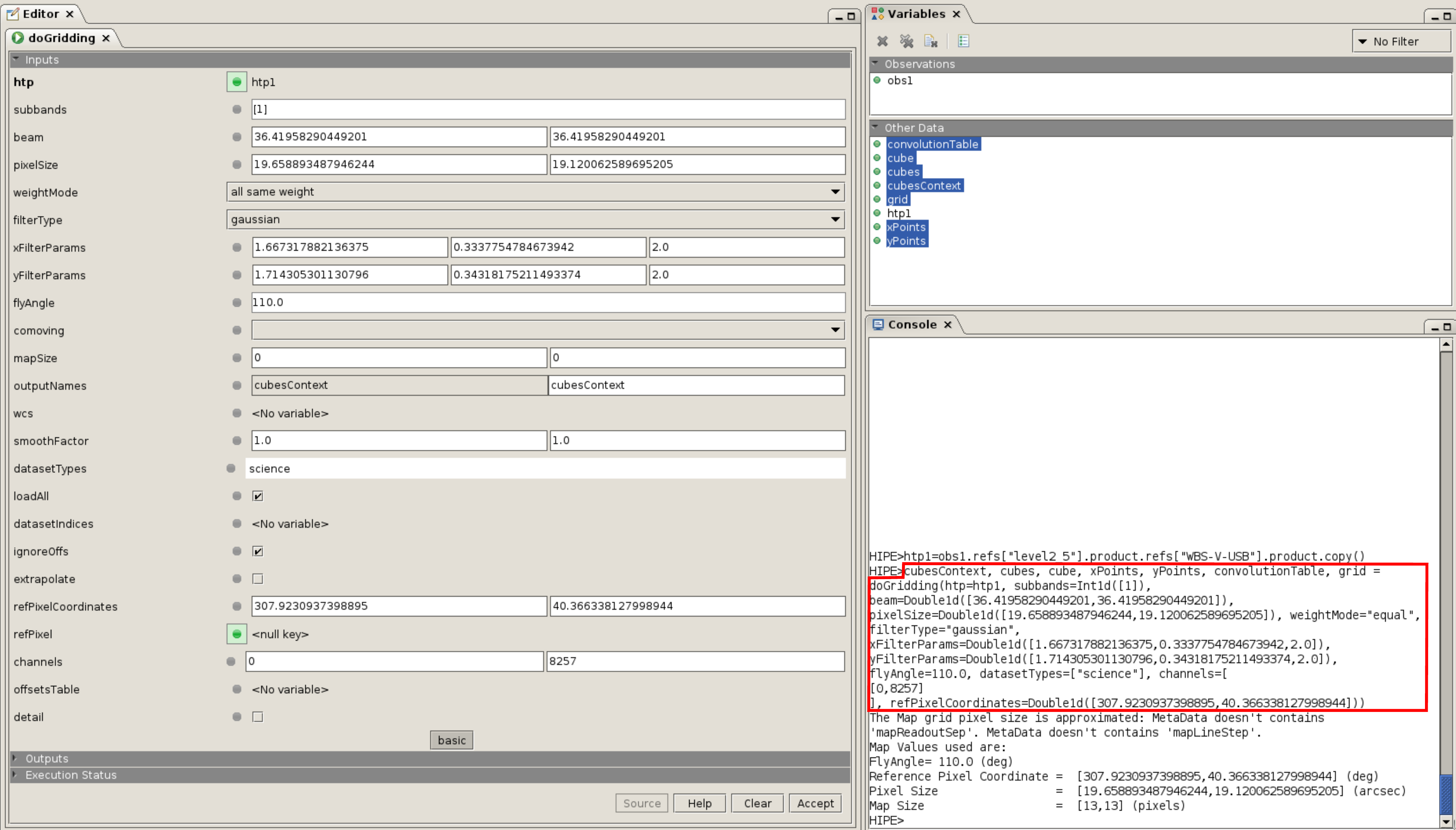}
   \caption{DoGriddingTask user interface (UI) including console output of the executed command. The top right panel displays the variables in HIPE. The variables highlighted in blue were created by executing the DoGriddingTask. The left panel displays the UI for the DoGriddingTask with several fields automatically generated when the HifiTimelineProduct, \textit{htp1}, was dragged from the variables section and dropped on the DoGriddingTask UI.  The bottom right panel shows the command line console with the red section showing the command that was generated when the DoGriddingTask was executed.}
   \label{fig:DoGriddingGUI}
   \end{center}
\end{figure*}

As this was the last level of the pipeline processing, there was value in summarizing the information contained in the observation to support more complex IA scenarios. The greater level of user interactions with observational data at this level of the pipeline required advanced Task GUIs in order to better manipulate the data.
The values displayed in the Task GUIs needed to be set to the default values, possibly dynamically determined from the data. 
Some complexity arose due to Task parameters that were interdependent, (e.g. changing the \textit{beam} values in the DoGriddingTask required changing the xFilterParameters and yFilterParameters values). Figure \ref{fig:DoGriddingGUI}, shows the user interface for the DoGriddingTask.  In this example, there were many parameters which could be adjusted, but not all of these were independent and the interface needed make sure that all parameters were self consistent (as well as provide a command line script, see section \ref{sec:processing}).

\subsubsection{Collecting summary information}

At the end of each pipeline level specific information about an observation was collected from the generated data products and provided to the user in the form of meta data, the most important information was collected and presented at the main ObservationContext level. Some summary parameters were an aggregation of parameters contained in the sub-product levels while others were calculated as a further pipeline step.

\begin{itemize}
\item UpdateObsMetaTask was executed at the end of each pipeline level. This task promoted specific sub-product information to the meta data of the ObservationContext. As this task was a simple aggregation of information, it was hidden from the end user and not included in the pipeline algorithm Jython scripts.

\item MkRmsTask was developed to compute a measure of RMS of the observational noise and used this value as a quality indicator of the observation.  The observed RMS value was compared directly to the predicted noise that had been calculated when planning the observation using the Herschel Observation Planning Tool.  As such, the comparison provided a measure of the success of the observation.

Before computing the RMS measure, the spectra were prepared by transforming them to the main beam temperature scale and for frequency switch modes, by folding the spectra \cite{HIFIH}. The task computed the RMS after identifying spectral features to avoid,  subtracting a baseline, and smoothing the spectra to a targeted resolution \cite{HIFIH}. The task was applied to each spectrometer as well after combining spectrometer polarizations.  For the mapping modes, the RMS was calculated per pointing position.  
\end{itemize}

\subsection{Post pipeline processing - Images and data quality}
\label{sec:browse}
\label{sec:quality}
The last stage of the pipeline created an overview and an assessment of the observation.  The overview was in the form of thumbnail images displaying the results of the observation and the assessment was in the form of a final quality consolidation of all quality information calculated during earlier pipeline steps.

The browse products and images provided a visual representation of the observation.  These were generated in the HIFI pipeline post processing plugin called the BrowsePlugin. 
This image was displayed with the observation when searching the HSA User Interface (HUI) for observations to download. 
They provided a \textit{quick-look} into the quality of the observation.  
This image could reveal issues such as baseline ripples that indicated the need for further corrections using interactive processing tools. 
The browse product was the basis for generating the browse image. 
There were three styles of browse images which could be generated by the plugin and these corresponded to the three main types of observing modes namely, Point, Mapping and Spectral Scan Mode.

The QualityContext was a data structure that contained all the generated quality information produced by the HIFI pipeline. 
The data structure was added by the QualityPlugin during the post-processing phase to the ObservationContext. The QualityPlugin would scan the entire processed observation, checking all the flags (a special type of meta data) raised from the executed pipeline tasks and summarized this information in the final QualityContext. 
Additionally, important information was collected from each quality container present at every level of the pipeline. These containers were a by-product of the data processing pipelines at each level. 

The possible flags were specified in a dedicated class and the QualityPlugin could identify and add them to the quality context. Utility classes were used to combine similar flags raised in different parts of the pipeline to create a suitable and compact overview on the issues found when processing an observation.

\section{Discussion} 
\label{sec:discussion}

The process of software development is shaped by the individuals within a project and the nature of the project. Many of the key individuals that influenced the early planning and design phase of the HCSS and the HIFI instrument pipeline were involved in previous satellite missions. They took their experiences, both good and bad, into this project with the goal of learning from those past experiences and improving the software developed for the Herschel mission. These lessons are summarized below as they shaped the software development process for the observatory and instrument teams.

\begin{itemize}
\item{{\bf Smooth transition}}

Contrary to earlier missions the Herschel software systems were designed and implemented in such a way that all mission phases used the same software for instrument commanding and data processing. This meant that the concepts for full operations had to be designed very early in the mission timeline even if the implementation would occur much later. This made it possible to process the data obtained during hardware testing with the \textit{same} software to be used during operations. This was called the smooth transition between testing and flight. This enabled the systems engineers to discover and solve complex hardware-software interaction problems much earlier than in previous missions. The users (such as instrument and calibration scientists) on one hand benefited from a uniform commanding and data analysis environment, but on the other hand had to cope with immature code at times.  

\item{\bf Interactive pipeline}

The Herschel mission required an automated pipeline to process all observations in a \textit{hands-free} manner. The instrument groups, responsible for the contents of the pipeline, needed to interactively modify the pipeline. This led to the concept of modes of operation for the pipeline as discussed in section \ref{sec:dataSoftware}. The pipeline processing was cut into small pieces, each one doing some well-defined part of the data reduction procedure. This gave flexibility to the instrument groups while supporting the needs of the Herschel mission.

It was hoped that much of the interactive analysis framework developed for Herschel data processing (e.g. HIPE) would be used in future projects. A considerable amount of effort developing and refining the HIPE user interface was centred on this belief, however without a proper plan in place to make this a reality meant that HIPE was a Herschel-only development.

\item{\bf Long lifetime}

Projects like Herschel have an exceptionally long lifetime. This project was started in the late 1990's and continued through 2017 at the end of post operations. Choices made at the beginning of the project had long lasting consequences. Many of the software tools used to support development at the beginning of the mission were very different at the end of the mission. Software developers started using Emacs or vi to write code and at the end of the mission were using Eclipse. Although git became available as a version control system during the the Hershel mission, the project continued to use CVS as the code repository and never migrated to git. The project developed an in-house build agent instead of using Jenkins and a custom bug tracking system before eventually migrating to Jira \textregistered. The project's choice to continue using CVS and our in-house build agent of instead upgrading had to be balanced against the disruption to the project versus the expected benefits. 


Many of the choices for software development tools served the project's needs well however a very small number of choices made resulted in additional unexpected overhead. This was particularly true of the choice to use a commercial object-oriented database to store the raw observational data. The original desire to have a strong connection between the observatory / instrument commanding data and the raw observational data lead to the decision to use a single object database. Object databases are better suited to applications where one needs to navigate through a complex series of objects (observatory / instrument commanding), however for observational telemetry we simply needed to make many queries based on the telemetry criteria. The observational telemetry is written once and accessed by many readers and this is better suited to an open source relational database system.

\item{\bf Platform independent and open source}

Another consequence of the long lifetime was that hardware changes were inevitable. At the beginning of the project, Sun\textsuperscript{TM} workstations and Windows\textsuperscript{TM} machines were widely used, this subsequently changed to Linux boxes and at the end of the project many people were working on Macbooks\textsuperscript{TM}. Though this particular succession of platforms could not be foreseen at the start of the mission, a succession of some kind was to be expected.  The choice of using the open source platform independent software Java\textsuperscript{TM} (combined later with Jython as interactive interface) above C++ was beneficial. It should be noted that choosing the Java\textsuperscript{TM} software language itself was risky at the time as it was only a couple of years older than the Herschel mission. On the other hand, the choice of Java\textsuperscript{TM} implied a huge investment into basic numeric libraries and in the development of astronomy processing code. 

Many of the issues relating to hardware and operating systems at the beginning of the mission have since been solved today with the advent of virtual machines and containerized environments allowing developers to develop for a given reference platform but run on almost any host operating system.

\item{\bf Software evolution}

The software developed within the project will need to evolve over the lifetime of the project, even the best designed software systems will still experience change. The HCSS software framework served the instrument teams well over the mission lifetime, but the instrument teams had to be prepared for change and have resources available to adapt to these changes. This was difficult to manage at times as the project tended to have the view that once a particular feature was code complete, the software would function as expected without change until the end of the mission. Upgrading the software reference platforms, Java\textsuperscript{TM} and Jython was necessary for operating system compatibility and security reasons. This was particular difficult in the case of Jython as these upgrades had a tendency to break existing code, requiring additional refactoring and testing. Whenever the system was not allowed to change, due to time or resource constraints, more effort was inevitably required by the users of the system. The lesson to draw from this is to plan for and accept that change is a normal part of iterative software development.


\item{\bf Target audience}

The target audience or \textit{end-user} of the software changed progressively over the lifetime of the Herschel mission.  In preflight, the HCSS software supported the instrument engineers and calibration scientists.  In this phase, the instrument engineers could directly communicate their requirements and issues with the software development team.  This process was efficient in resolving issues and providing the needed capabilities. After launch, a whole new group of users began to use the software, namely the broader astronomical community.  The communication between the software development team and this broader group of users was not as seamless as with the instrument scientists.  
It is important to recognize when the target audience for the software has changed and the additional requirements the new audience brings to the project.

\item{\bf Software development practices}

The Herschel project was a collaboration of independent institutes across Asia, Europe and North America. It was hard to enforce code quality standards on contributions across packages with multiple independent institutes providing contributions. HIFI software development team found that increasing the frequency of in-person working sessions (co-locations) where the developers met in a single location for upwards of three weeks every four months helped to mitigate the effects of the distributed team and the varied skill level of the developers. Implementing additional processes including paired programming and code reviews would have helped to increase the quality of the code and better train developers just joining a project. 
These investments can be expensive initially, however over the lifetime of the project, they will reduce maintenance costs by minimizing time wasted checking and fixing coding bugs.

The public release of the Herschel software was largely time based. The project would select some date in the future and then schedule development work around a particular release date. Software release candidates were branched from the main development track and bug fixes were applied to the release branch until all critical issues were resolved. Software testing was performed by staff astronomers to ensure that the results generated by the software were correct and the software supported the necessary work-flows required to process and analyze the observational data. The development builds were publicly available although they were only intended for use by the developers and the instrument staff scientists. The reduction in the amount of time between the development and release of new features or bug fixes is important to increase usefulness and acceptance of the software. The DevOps model of software development and deployment encapsulates this practice.

\item{\bf Research based software projects}

Research based software development projects, space science in this case, are different than developing software for commercial purposes. It is not always clear or completely understood what is needed at the onset and many prototypes are likely to be developed to test behaviour before the overall design is finalized. This occurred within the HIFI instrument team leading to more prototype like code being used as production code at times, with the unit and system testing code being written later in the project. When the target audience changes occur in these cases, it is important to stay focused on delivering a quality product with proper software testing support in order to minimize the transition costs and increase the acceptance of the product by the new audience.

Given the specialized nature of the problem space, it is important to have a combination of astronomers, system engineers and software engineers working in close proximity to each other to produce the best possible outcome for the project. Institutes participating in projects like these need to recognize that developing software requires a significant amount of resources to be done properly.
 
\end{itemize}

\section{Conclusions}
\label{sec:summary}

The HIFI pipeline was designed to fit into a common software infrastructure and to be executed in four 
distinct modes (e.g., lab, interactive, systematic and on-demand). The software development process was supported by a set of tools and processes to ensure the goals of building a robust and efficient pipeline were accomplished. The software build and test environment was developed to help monitor the changes to the pipeline and quickly identify undesired changes so that issues could be resolved efficiently. 

The main accomplishment of the HIFI pipeline was creating a single pipeline for all HIFI observations 
regardless of the observing mode and type of observation taken by the instrument.  By using a single 
pipeline for all modes code duplication was reduced ensuring consistency across generated data 
products. With the customizable pipeline, HIFI was able to support robust interactive data analysis work-flows with a small amount of additional complexity in the pipeline software. 


The HIFI pipeline software, supporting auxiliary and calibration data products were the end result of 
accomplishing the goal of providing high quality observational data products to the scientific community. The result of this work is a set of data products contained in the HSA that have been used to do ground breaking research resulting a large number of published scientific papers in multiple refereed scientific journals. As of 4 January  2019, 4403 out of 8571 HIFI observations have appeared in refereed journals. 
Given the long lead time and duration of the Herschel/HIFI project many of the initial developments have been superseded in the software industry, however the general approach remains pertinent  even today.  
The content of this article should serve as a \textit{lessons-learned} for future projects that are considering developing their own data processing software infrastructure.

\section*{Acknowledgements}
{HIFI was designed and built by a consortium of institutes and university departments from across
Europe, Canada and the United States under the leadership of SRON Netherlands Institute for Space 
Research, Groningen, The Netherlands, and with major contributions from Germany, France and the US.
Consortium members are: Canada: CSA, University of Waterloo; France: CESR, LAB, LERMA, IRAM; 
Germany: KOSMA, MPIfR, MPS; Ireland, NUI Maynooth; Italy: ASI, IFSI-INAF, Osservatorio Astrofisico 
di Arcetri-INAF; Netherlands: SRON, TUD; Poland: CAMK, CBK; Spain: Observatorio Astron Nacional 
(IGN), Centro de Astrobiologia (CSIC-INTA). Sweden: Chalmers University of Technology - MC2, RSS \& 
GARD; Onsala Space Observatory; Swedish National Space Board, Stockholm University - Stockholm 
Observatory; Switzerland: ETH Zurich, FHNW; USA: Caltech, JPL, NHSC.}

\bibliographystyle{elsarticle-num}
\bibliography{./hifi.bib} 

\end{document}